\documentstyle[aps,pre,multicol,epsf,epsfig]{revtex} 
\begin{document}
 \def\B.#1{{\bbox{#1}}}
\title{{\rm PHYSICAL REWIEV E \hfill Sumbitted}\\
On Conditional Statistics in Scalar
  Turbulence: Theory vs. Experiment} 
\author{Emily S.C. Ching$^*$,
  Victor S. L'vov$^{\dag\ddag}$, Evgeni Podivilov$^{\ddag}$, and
  Itamar Procaccia$^{\dag}$} 
\address{$^*$Department of Physics, The
  Chinese University of Hong
  Kong, Shatin, Hong Kong, \\
  $^{\dag}$ Department of Chemical Physics, 
The Weizmann Institute of Science, Rehovot 76100, Israel\\
  $^{\ddag}$ Institute of Automation and Electrometry, Ac. Sci. of
  Russia, 630090, Novosibirsk, Russia} 
\maketitle 
\begin{abstract}
  We consider turbulent advection of a scalar field $T(\B.r)$, passive
  or active, and focus on the statistics of gradient fields
  conditioned on scalar differences $\Delta T(R)$ across a scale $R$.
  In particular we focus on two conditional averages $\langle\nabla^2
  T\big|\Delta T(R)\rangle$ and $\langle|\nabla T|^2\big|\Delta
  T(R)\rangle$. We find exact relations between these averages, and
  with the help of the fusion rules we propose a general
  representation for these objects in terms of the probability density
  function $P(\Delta T,R)$ of $\Delta T(R)$. These results offer a new
  way to analyze experimental data that is presented in this paper.
  The main question that we ask is whether the conditional average
  $\langle\nabla^2 T\big|\Delta T(R)\rangle$ is linear in $\Delta T$.
  We show that there exists a dimensionless parameter which governs
  the deviation from linearity. The data analysis indicates that this
  parameter is very small for passive scalar advection, and is
  generally a decreasing function of the Rayleigh number for the
  convection data. 
\end{abstract} 
\begin{multicols}{2}\narrowtext
\section{Introduction}
The equations of motion in fluid mechanics, be them for the velocity
field $\B.u(\B.r,t)$ or for a scalar field like the temperature
$T(\B.r,t)$, contain interaction terms like $\B.u\cdot\B.\nabla \B.u$
or $\B.u\cdot\B.\nabla T$ and dissipative terms like $\nu \nabla^2
\B.u$ or $\kappa \nabla^2 T$, with $\nu$ and $\kappa$ being the
kinematic viscosity and the scalar diffusivity respectively.
Accordingly, when one attempts to derive a theory of correlations
functions of the field or of field differences across a length scale
$R$ one runs into mixed correlation functions of the Laplacian of the
field with the field itself. At this point there are two fundamentally
different approaches that we want to expose first. We exemplify these
approaches in the context of turbulent advection, but similar
considerations
apply also to Navier-Stokes turbulence.

One fundamental strategy, which is the more usual one, is to consider
the structure functions of the field differences. Denoting $\Delta
T(\B.r,\B.r')\equiv T(\B.r')-T(\B.r)$, the structure functions
$S_n(R)$ are defined by
\begin{equation}
S_n(R) = \left<[\Delta T(\B.r,\B.r')]^n\right>,
 \quad R=|\B.R|\equiv |\B.r'-\B.r| \ , \label{Sn}
\end{equation}
where homogeneity and isotropy of the ensemble were assumed. Using the
advection equation
\begin{equation}
{\partial T\over \partial t} +\B.u\cdot\B.\nabla T
 = \kappa \nabla^2 T \label{eq}
\end{equation}
one can derive the equation of motion for $S_n(R)$: 
\begin{eqnarray}
  {\partial S_n(R)\over \partial t} &+&D_n(R) =J_n(R) \ , \label{EqSn}\\
D_n(R) &=& 2n \langle [\B.u(\B.r)\cdot \B.\nabla 
\Delta T(\B.r)] [\Delta T(\B.r,\B.r')]^{n-1}\rangle\ , 
\label{defDn}\\ J_{n}(R) &=& -2n\kappa \langle 
\nabla^2 T(\B.r)[\Delta T(\B.r,\B.r')]^{n-1} \rangle\ . \label{J2n}
\end{eqnarray}
In the stationary state we have a ``balance equation" $D_n(R)=J_n(R)$. 

The obvious advantage of this approach is that it involves objects
that depend on one coordinate $R$ only. On the other hand the analysis
of the balance equation requires a theory of the {\em viscous} term
$J_n$ which involves a correlation between the Laplacian of the field
and field differences. Even when $R$ is within the inertial range, one
cannot get rid of $J_n(R)$. The limit $\kappa \to 0$ does not help;
$J_n$ has a finite value in this limit.

A second fundamental approach which avoids this difficulty
\cite{96LP-1} employs multi-point correlation function of field
differences. Starting form the same field differences $\Delta
T(\B.r_1,\B.r'_1)$ one defines the correlation function
\begin{equation} {\cal F}_n(\B.r_1,\B.r'_1,\dots \B.r_n,\B.r'_n)
  \equiv \left<\Delta T(\B.r_1,\B.r'_1)\dots \Delta
    T(\B.r_n,\B.r'_n)\right> \ , \label{defFn} \end{equation} which
depends on $n$ fields and $2n$ coordinates. The equation of motion for
${\cal F}_n$ looks superficially similar to (\ref{EqSn}):
\begin{equation} {\partial {\cal F}_n\over \partial t} + {\cal
    D}_n(\B.r_1,\B.r'_1,\dots \B.r_n,\B.r'_n) ={\cal
    J}_n(\B.r_1,\B.r'_1,\dots \B.r_n,\B.r'_n) \ , \label{balm}
\end{equation} where
\begin{eqnarray}
&& \B.{\cal D}_n(\B.r_1,\B.r'_1;\dots
\B.r_n,\B.r'_n)=\sum_{j=1}^n\langle\Delta 
T(\B.r_1,\B.r'_1)\dots [\B.u(\B.r'_j)
\cdot\B.\nabla_{j'}T(\B.r'_j)\nonumber \\ &&-\B.u(\B.r_j)
\cdot\B.\nabla_{j}T(\B.r_ j)] \dots \Delta T(\B.r_n,\B.r'_n)\rangle. 
\label{Dnm}\\ && \B.{\cal J}_n(\B.r_1,\B.r'_1;\dots
\B.r_n,\B.r'_n)=\kappa\sum_{j=1}^n\left(\nabla_j^2+\nabla_{j'}^2\right)
 \langle\Delta T(\B.r_1,\B.r'_1)\dots \nonumber \\
 && \dots \Delta T(\B.r_j,\B.r'_j)
\dots \Delta T(\B.r_n,\B.r'_n)\rangle \ . \label{Jnm}
 \end{eqnarray}
 In the stationary state we again face a ``generalized balance
 equation" ${\cal D}_n={\cal J}_n$. In fact, there is a fundamental
 difference between the two approaches. In the present case one can
 analyze the generalized balance equation keeping all the separations
 within the inertial interval of scales. Then we can take the limit
 $\kappa \to 0$ and show \cite{96LP-1} that the dissipative term
 ${\cal J}_n$ vanishes in the limit. We remain in this limit with a
 homogeneous equation ${\cal D}_n=0$ without any Laplacian terms. The
 advantage is that we have in principle a complete theory without the
 need of additional input. The obvious disadvantage of this approach
 is that we have functions of many variables. Nevertheless, this
 approach turned out to be very useful in the context of Kraichnan's
 model for passive scalar advection \cite{68Kra}, where the
 homogeneous equation can be turned into a linear partial differential
 equation for the correlation functions ${\cal F}_n$. But even in this
 simplest possible case the difficulty incurred by having functions of
 many variables led to contradicting arguments about the relevant
 physical solution.
 
 It is thus obviously useful to find ways to analyze the simpler
 version in which we have one variable only, but a mixture of inertial
 and dissipative scales contributing to the correlation function
 $J_n$. To proceed we need however additional information. One
 possible way of inputting this information is in the language of
 conditional averages. To see this consider the mean value of the
 $\kappa\nabla^2 T$ condition on $\Delta T$:
\begin{equation}
H(\Delta T,R)\equiv \kappa\left<
\nabla^2 T(\B.r)\Big|\Delta T(\B.r,\B.r')\right> \ . \label{defH}
\end{equation}
Using this definition we rewrite $J_n(R)$ as \begin{equation}
\label{Jn2}
J_{n}(R)=-2n\int d\Delta T P(\Delta T,R) 
[\Delta T]^{n-1} H(\Delta T,R) \ , \label{int1} 
\end{equation}
where $P(\Delta T,R)$ is the probability density function (pdf) to
find a temperature difference $\Delta T$ across a separation $R$. It
was proposed by Kraichnan \cite{94Kra} that the conditional average
$H(\Delta T,R)$ exhibits in his model a very simple functional
dependence on $\Delta T$ and on $R$, i.e.
\begin{equation}
H(\Delta T,R) = {-J_2 \Delta T(R)\over 4 S_2(R)} \ . \label{Kra} 
\end{equation}
If this were true, $J_n(R)$ could immediately be written in terms of
structure functions,
\begin{equation}
J_n(R) = {n J_2 S_n(R) \over 2 S_2(R)} \ . \label{fuse} 
\end{equation}
This again closes the theory upon itself, allowing one to proceed. In
the Kraichnan model, $D_n$ can also be written in terms of $S_n(R)$
[see Eq.(\ref{exa}) below] and one can therefore find the scaling
exponents that characterize the structure functions. Unfortunately
there is still no derivation of the ansatz (\ref{Kra}). There are
indications that it is obeyed; numerical simulations of the Kraichnan
model support it rather convincingly \cite{95KYC}. In addition it was
shown in \cite{96CLP}, on the basis of experimental data analysis,
that this form of the conditional average is obeyed in a context that
is much wider than the Kraichnan model.  Experiments on turbulent
advection were analyzed, and good agreement with (\ref{Kra}) was
demonstrated. The aim of this paper is to present further theoretical
and data analysis in this direction. We want to understand what can be
said on conditional averages in terms of fundamental theory, and how
to intelligently analyze experimental data to probe these important
quantities.

It should be pointed out that although we focus in this paper on
turbulent advection, similar considerations are important also in the
context of Navier-Stokes turbulence. Also in that case the two
fundamental strategies to develop a statistical theory are open to us.
The second strategy is even more tempting in that context. In the
first approach one gets objects depending on one coordinate, but a
hierarchy of equations relating different orders (in powers of the
velocity field) correlation functions. The second strategy gives a
theory involving many coordinates, but in which we can also neglect
the viscous term, obtaining homogeneous equations ${\cal D}_n=0$ that
involve only one order of correlation functions. We are not going to
explore this issue further and refer the reader to \cite{96LP-1} for
more details.

The paper is organized as follows: In Secion II we present theoretical
considerations that relate conditional averages with the probability
density function. The fusion rules are employed to develop a general
representation of the conditional average (\ref{defH}) in terms of the
pdf of $\Delta T(R)$. It is shown that
in general $H(\Delta T,R)$ can be written as an expansion in
non-integer powers of $\Delta T$, with the first term being linear,
and with dimensionless coefficients that are denoted $a_0, a_1\dots$,
see Eqs.(\ref{H1}),(\ref{opL}). In Section III we analyze experimental
data of passive and active scalar advection, with the aim of
understanding whether the linear term in our expansion is leading, and
whether the rest of the series is unnecessary. We offer conclusion in
Section IV: it turns out that for passive scalar advection the linear
fits are excellent, whereas in the case of active convection the
linear form appears to fit the data extremely well for high values of
the Rayleigh numbers Ra. For lower values of Ra there are significant
nonlinear contributions, and we show that the proposed method of data
analysis offers excellent fits to the data.
\section{Conditional Averages and the Relations Between Them} 
\subsection{The conditional average of the dissipation field}
In addition to the conditional average (\ref{defH}) we will conisder
the average of the scalar dissipation field $\kappa |\B.\nabla T|^2$
conditioned on $\Delta T$:
\begin{equation}
G(\Delta T,R)=\kappa\left<|\B.\nabla T|^2 
\Big | \Delta T(\B.r,\B.r')\right> \ . \label{defG}
 \end{equation}
 This conditional average appears naturally in the analysis of
 $J_n(R)$. For space homogeneous statistics one can move one gradient
 around in the definition (\ref{J2n}) and get for $R$ in the inertial
 range,
\begin{equation}
J_{n}(R) \sim -2n(n-1)\kappa \langle |\B.\nabla 
T(\B.r)|^2 [\Delta T(\B.r,\B.r')]^{n-2}
\rangle \ . \label{secondJ}
\end{equation}
Accordingly, we can write a second equation in terms of the
probability density function
\begin{equation}
\label{Jn3}
J_{n}(R)=- 2n(n-1) \int d\Delta T P(\Delta T,R) 
[\Delta T]^{n-2} G(\Delta T,R) \ .
\end{equation}
By equating (\ref{Jn3}) with (\ref{Jn2}) we find an infinite set of
integral constraints on the conditional averages. This implies that
the two conditional averages, $H$ and $G$, must be universally related
in order to satisfy these constraints for any value of $n$. The
required relationship involves the pdf $P(\Delta T,R)$ and has the
following form:
\begin{equation}
\label{univ}
H(\Delta T, R) P(\Delta T,R) = 
- {\partial \over \partial \Delta T} \left[ G(\Delta T, R) P(\Delta T,
R)
 \right] \ ,
 \end{equation}
 as can be checked by direct substitution. A formula of this form has
 been discussed before in \cite{96Ching}.
\subsection{Fusion rules and their consequences}
An additional constraint on the conditional averages can be obtained
using the ``fusion rules" that have been derived recently. These rules
serve to find relationships between the two fundamental approaches
described in the introduction. Specifically, the fusion rules address
the asymptotic properties of ${\cal F}_{n}$ when a group of $p$
points, $p<n-1$ tend towards a common point $\B.r_0$ $(\left|
  \B.r_i-\B.r_0\right|\sim \rho$ for all $i\le p$), while all the
other coordinates remain at a larger distance $R$ from $\B.r_0$
$(\left|\B.r_i-\B.r_0\right|\sim R$ for $i>p$, and $R\gg \rho$). For
our particular purposes we need to write $J_n(R)$ as the result of the
following fusion process:
\begin{equation}
J_n(R) = -2n\kappa \lim_{\B.r_i\to \B.r_0} \lim_{\B.r'_i\to 
\B.r_0+\B.R} \B.\nabla^2_{r_1} {\cal F}_n(\B.r_1,\B.r'_1\dots
 \B.r_n,\B.r'_n)
 \end{equation}
 The fusion rules that should be used in such cases were displayed in
 great detail in \cite{96LP-1} in the context of Navier-Stokes
 turbulence. They apply identically also to this case. Basically it
 was shown that all the fusions without gradients in this case have
 regular limits, relating ${\cal F}_n$ with $S_n$. The fusions with
 gradients require special care of the limit $\B.r_{12}\equiv
 \B.r_1-\B.r_2 \to 0$. The intermediate result is, for $R$ in the
 inertial range, \begin{equation} J_n(R) \sim -2n\kappa
   \lim_{\B.r_{12}\to 0} \B.\nabla^2_{r_1}S_2(r_{12}) S_n(R)/S_2(R) \ 
   . \label{inter}
\end{equation}
In evaluating this quantity we interpret the limit $\B.r_{12} \to 0$
as a limit $r_{12}\to \eta$. This seems natural for large Peclet
numbers when $\eta \to 0$. It is important however to stress that
there is a hidden assumption here. We expect the function ${\cal
  F}_{2n}$ to change its analytic behavior as a function of $r_{12}$.
This change occurs at the viscous crossover scale $\eta$. The issue is
whether this crossover scale is $n$ and $R$ independent. That this is
so has been {\em proven} for Kraichnan's model of turbulent advection
\cite{96LP}, and that this is {\em not} so has been proven for
Navier-Stokes turbulence \cite{96LP-1}. We believe that this is more
generally true in scalar advection due to the linearity of the
equation of motion (\ref{eq}), independently of the statistical
properties of the driving velocity field. The experimental results
analyzed in \cite{96CLP} strongly indicate that this is the case in a
wide context of turbulent scalar fields. With this in mind we write
\begin{equation} J_{n}(R) \sim -2\kappa n
  \big[\nabla^2_{r_{12}}S_2(r_{12}) |_{r_{12}=\eta}\big]
  S_{n}(R)/S_2(R)\ . \ 
\end{equation}
Using the fact that the mean of the scalar dissipation field, denoted
$\bar\epsilon$, is evaluated as $\bar\epsilon \sim
\kappa\big[\nabla^2_{\rho}S_2(\rho)|_{\rho=\eta}\big]$, and also the
fact that in the inertial range $J_2(R)=-4\bar\epsilon$, we write
\begin{equation} J_{n}(R) = {n C_{n} J_2 S_{n}(R) \over 2 S_2(R)}\ ,
  \label{almostfinal}
 \end{equation}
 where $C_{n}$ is an as yet unknown dimensionless coefficient, but
 $C_2=1$. Eq.(\ref{almostfinal}) was suggested for Kraichnan's model
 in \cite{94Kra} and derived in \cite{96FGLP}. We proposed in
 \cite{96CLP} that it holds in a much wider context and showed
 experimental data in support.
 
 Having the result (\ref{almostfinal}) we see that the scaling
 exponent of $J_n(R)$ is fixed as $\zeta_n-\zeta_2$. One way to
 understand it is to assume that indeed Eq.(\ref{Kra}) is always
 valid, and to use it in Eq.(\ref{int1}) to derive this scaling law.
 In this case Eq.(\ref{almostfinal}) is recovered with the constraint
 that the coefficients $C_n$ are all unity and in particular
 $n$-independent. We need however to allow for the possibility that
 (\ref{Kra}) is incorrect, and find alternative representations of the
 conditional average that agree with the scaling law
 (\ref{almostfinal}). This is the subject of the next subsection.
\subsection{Series expansion of the conditional average}
Let us reject for the time being the possibility that $H(\Delta T,R)$
is proportional to $\Delta T(R)$ with a coefficient depending only on
$R$. Alternatively, let us consider the following model expression for
$H(\Delta T,R)$:
\begin{equation}
\label{H1}
H(\Delta T,R)= {-J_2\over 4 S_2(R)}\,
 { \hat{\cal L}\, \big\{ \Delta T \,P(\Delta T,R)\big\}
 \over P(\Delta T,R)} . 
\end{equation}
Here we introduce the dimensionless operator $\hat{\cal L}$ acting on
the variable $\Delta T(R)$ as a sum of differential operators:
\begin{equation}
\label{opL}
\hat{\cal L}= \sum_{p=0}^\infty {a_p \over p!} 
\Big[{\partial \over \partial \Delta T }\Big]^p (\Delta T)^p \ .
\end{equation}
In this representation there is the freedom of a countable set of
dimensionless coefficients $a_p$.

From the dimensional point of view $ H(\Delta T,R)$ in (\ref{H1}) is
of the order of $\Delta T$, but it has a more complicated functional
dependence on $\Delta T$ and $R$, expressed in terms of the pdf
$P(\Delta T,R)$ with the help of operator $\hat{\cal L}$. Computing
$J_{n}(r)$ with $H$ given by (\ref{H1}), one gets
Eq.(\ref{almostfinal}) as we should, but with coefficients $C_n$ given
by
\begin{equation}
\label{Cn}
C_n= \sum_{p=0}^{n-1} \Big(\begin{array}{c}n-1\\ 
p\end{array}\Big)(-1)^p a_p\ .
\end{equation}
Here
$$\Big(\begin{array}{c}n-1\\ p\end{array}\Big) = {(n-1)!\over p!
  (n-p-1)!}$$
are binomial coefficients. We have one obvious
constraint, i.e. $C_2 = a_0 - a_1 = 1$. One sees that by an appropriate
choice of $a_p$, an arbitrary dependence of $C_n$ on $n$ is possible.

To exemplify the consequences of this extra freedom we will analyze
next the implications it has on the scaling exponents of the Kraichnan
model of passive scalar convected by an infinitely fast Gaussian
velocity field. In this case the term $D_n(R)$ is known exactly,
\begin{equation} D_n(R)={{\cal D}\over R^{d-1}}{d\over dR}
  R^{d-1+\zeta_h}{d\over dR}S_n(R) \ , \label{exa}
\end{equation}
where $\zeta_h$ is the parameter of the model, $0<\zeta_h<2$, and
 $\cal D$ is a dimensional constant. Using the balance equation and
writing $S_n(R)\propto R^{\zeta_n}$ one computes
\begin{equation}
\label{zetan}
\zeta_n=\case{1}{2}\sqrt{(d-2)^2 + 2nd\zeta_2 C_n}-\case{1}{2}(d-2).
 \end{equation}
 If all coefficients $a_{(m\geq 1)}$ are zero, then $C_n=1$ and we
 have Kraichnan's conjecture for $\zeta_n$:
\begin{equation}
\label{zetan0}
\zeta_n=\case{1}{2}\sqrt{(d-2)^2 + 2nd\zeta_2}
-\case{1}{2}(d-2) 
\end{equation}
which dictates the ``square-root'' asymptotic behavior ~$\zeta_n
\to\sqrt{nd\zeta_2/2}$ in the limit $n\to \infty$. Assume now that  
only $a_0$ and $a_1$ are
nonzero. From (\ref{Cn}), $C_n = a_0 - (n-1) a_1$ or
 \begin{equation}
  C_n=1-(n-2)a_1\ , \label{Cneq}
\end{equation}
and if we use this result in the Kraichnan model we get
\begin{equation}
\label{zetan1}
\zeta_n=\case{1}{2}\sqrt{(d-2)^2 + 2nd\zeta_2[1-(n-2)a_1]}
 -\case{1}{2}(d-2)\ . 
\end{equation}
Now the asymptotics of $\zeta_n$ in $n$ are linear: $\zeta_n\propto
n\sqrt{-a_1}$ for $n \to \infty$.  Notice that in this case, $a_1$ has
to be negative. It is interesting to note that the assumption that
only the first three coefficients, $a_0$, $a_1$, and $a_2$ are nonzero
would lead to the conclusion that $\zeta_n \propto n^{3/2}$, which is
not allowed in view of the Hoelder inequalities for the scaling
exponents. Similarly, one cannot truncate the series (\ref{opL}) at
any higher term. Hence, only three possibilities are allowed for this
representation of the conditional average: 
\begin{itemize}
\item[(i)] only $a_0$ is nonzero and we have Kraichnan's exponents
  (\ref{zetan0}); In this case we expect to find a {\em linear} law
  \begin{equation} H(\Delta T,R)={-J_2\over 4 S_2(R)}\Delta T(R)\ .
    \label{linear} 
\end{equation}
\item[(ii)] only $a_0$ and $a_1$ are nonzero and we have the exponents
  (\ref{zetan1}); Note that Eq.(\ref{Cneq}) determines the
  coefficients $C_n$ in this case, and the nagnitude of the
  dimensionless parameter $a_1$ measures the deviation of $C_n$
  from unity. We will see below that in all high Re data $a_1$ seems
  to be smaller than $10^{-2}$.
 \item[(iii)] there is an infinite set
  of non-zero coefficients $a_p$. 
\end{itemize}

It is interesting to ask whether one can come up with an example of
infintely many coefficients $a_p$ without violating any general
requirement about the scaling exponents. In fact this can be easily
done. For example, choose $a_p$ of the following form 
\begin{equation}
  a_p = \sum_s \beta_s [1-\exp(-\alpha_s)]^p + \delta_{p0}\mu -
  \delta_{p1}\nu\ ,
\end{equation}
and substitute in Eq.(\ref{Cn}). The result is 
\begin{eqnarray}
\nonumber
&&C_n = \mu + (n-1)\nu \\
&+& \sum_{p=0}^{n-1}\Big(
\begin{array}{c}n-1\\ p\end{array}\Big)
\sum_s \beta_s (e^{-\alpha_s}-1)^p
\label{exp}
\end{eqnarray}
Finally it gives:
\begin{equation}
C_n=\mu + (n-1)\nu +\sum_s \beta_s \exp[-(n-1)\alpha_s]\ , 
\end{equation}
 wich satisfies the constraint
\begin{equation}
C_2 =\mu +\nu + \sum_s \beta_s \exp(-\alpha_s)=1\ . 
\end{equation}
In the limit $n \gg \max [1/\alpha_s]$ we find 
\begin{equation}
\label{zetalimit}
\zeta_m=\case{1}{2}\sqrt{(d-2)^2 + 2nd\zeta_2[\mu+(n-1)\nu]} 
-\case{1}{2}(d-2)\ .
\end{equation}
This form has again an asymptotic linear dependence of $\zeta_n$ on
$n$, but for intermediate values of $n$ these exponents differ
significantly from (\ref{zetan1}). We do not ascribe particular
importance to this result, and exhibit only to show that to satisfy
the consequences of the fusion rules, we have in general considerable
freedom in the functional dependence of $\zeta_n$ on $n$.

It is important to understand now that the series (\ref{H1}) is
actually an expansion in {\sl non-integer powers} of $\Delta T$. As
such, it is fundamentally different from the series proposed in
ref.\cite{95KYC} which is in {\sl integer} powers. The non-integer
powers are dictated by the functional form of the pdf $P(\Delta T,R)$,
which in general is non-analytic. In order to see this clearly, we
consider for example a form $P(\Delta T,R)$ that has been found
\cite{91Ching} to fit very well the experimental data for turbulent
temperature fluctuations. For different separations $R$, the pdf is
described by the following stretched-exponential form:
\begin{equation}
\label{pdf}
P(\Delta T, R) =
C(R)\exp\left[-\alpha(R)|\Delta T|^{\beta(R)} \right] \ .\label{pdefex}
 \end{equation}
 Substitution of this into Eq. (\ref{H1}) gives a series in {\sl non
   integer} powers of $\Delta T$ which originate from the
 differentiation of $(\Delta T)^\beta$. Any attempt to re-expand the
 series in $(\Delta T)^{m \beta}$ for {\sl non-integer $\beta$} in
 {\sl integer} powers of $\Delta T$ leads unavoidably to a series with
 zero radius of convergence.
\section{Is the conditional average $H(\Delta T,R)$ linear or nonlinear
    in $\Delta T$?} 
  
  As we already saw, the present state of the theory does not allow
  an ab-initio determination of the functional dependence of the
  conditional average $H$ on $\Delta T$. Accordingly, we turn now to
  analyzing experimental data to shed light on this issue. As
  explained, the conditional average $H(\Delta T,R)$ is linear in
  $\Delta T$ if and only if all the coefficients except $a_0$ are zero
  (i.e. the possibility (i) discussed in last section). For the other
  two cases, $H(\Delta T,R)$ is a nonlinear function of $\Delta T$. In
  earlier work\cite{96CLP}, we found that $H(\Delta T,R)$ is close to
  a linear function of $\Delta T$ which implies that $a_p, p\ne0$ are
  small compared to $a_0$. To make more quantitative statements we
  will perform further analysis of the experimental data under the
  assumption that $a_0$ and $a_1$ are nonzero. We will see that this
  form fits the data extremely well, and that the coefficient $a_1$ is
  always small, and it appears to become smaller when the Reynolds
  number is increasing and when the chosen separation goes into the
  bulk of the inertial interval. Taking $a_0$ and $a_1$ as the only
  nonzero coefficients we find 
\begin{eqnarray} 
\nonumber
    && H(\Delta T,R) \\
&=&
{-J_2 \Delta T \over 4 S_2(R)} 
\left\{ (a_0+2a_1) + a_1 \Delta T {\partial 
\left[ \log P(\Delta T,R) \right]
\over \partial \Delta T}
\right\}.
\end{eqnarray}
Thus, the coefficient $a_1$ indeed measures how nonlinear $H$ is.
Using the form (\ref{pdefex}) for the pdf $P(\Delta T,R)$, $H$ can be
rewritten as
\begin{eqnarray}
\nonumber
&& H(\Delta T,R) \\
&=& {-J_2 \Delta T \over 4 S_2(R)} \left[ (1+3a_1) - a_1 
\alpha(R) \beta(R) |\Delta T|^{\beta(R)} \right] 
\label{Hform}
 \end{eqnarray}
 in which $a_0=1+a_1$ is used. When $P(\Delta T,R)$ is asymmetric, a
 more general form with different $\alpha$ and $\beta$ for $\Delta T >
 0$ and $\Delta T < 0$ has to be used. Note that if we measure $\Delta
 T$ in units of its standard deviation (as we do below in the data
 analysis), then the combination of parameters $a_1 \alpha(R)
 \beta(R)$ gives a direct measure of the importance of the nonlinear
 correction in this equation for $\Delta T =1$.
\subsection{Linear Fits}
We begin the discussion of experimental data by demonstrating that in
the case of passive scalar advection the linear form of the
conditional average $ H(\Delta T,R)$ is observed to high precision.
Firstly we examine the theoretical prediction (\ref{almostfinal}). The
results show that to a good accuracy $C_{n}\approx 1$ for all $n$ and
$R$\cite{96CLP}.
\begin{figure}
\epsfxsize=7.8truecm
\caption{ A plot of $\log|J_{2n}(R)/(2n\kappa)|$ vs. 
$\log|(2\kappa)^{-1}J_2 S_{2n}(R)/ S_2(R)|$ for $n=2$~(squares),
  3~(triangles), 4~(diamonds), 5~(stars), and 6~(circles) and $R$
 in the inertial range. The data are taken from Yale [10]. The line
 (slope 1 and intercept 0) is not a fit but is the theoretical 
expectation (21) with $C_n =1$.} \label{Fig1}
\end{figure}


We use temperature data measured in the wake of a heated cylinder
\cite{Sre}. Air of speed 5 m/s flowed past a heated cylinder of
diameter 19 mm (Reynolds number $=9.5\times10^4$). The temperature was
measured at a fixed point downstream of the cylinder on the wake
centerline. The cylinder was heated so slightly that the buoyancy term
was unimportant and the temperature acted as a passive scalar.
Temperature
was measured as a function of time, and we use here the standard
Taylor hypothesis that surrogates time derivatives for space
derivatives. In doing so we made sure that the viscous scales are
properly resolved in this data set. In Fig.~1 we display
$J_{2n}(R)/(2n\kappa)$ as a function of
$(2\kappa)^{-1}J_2S_{2n}(R)/S_2(R)$ for $n$ varying from 2 to 6, and
for various $R$ values in the inertial range. We see that all the
points fall on a line whose slope is unity to high accuracy, and whose
intercept (in log-log plot) is very closely zero. As was pointed out
in \cite{96CLP} this good agreement is a confirmation of the validity
of the fusion rules. It should be stressed that individual tests at
various values of $n$ as a function of $R$ corroborate the same
conclusion, i.e. Eq.(\ref{almostfinal}) is supported by the
experimental data with $C_{2n}$ being near unity. The most sensitive
test of the alleged constancy of the coefficients $C_{2n}$ is obtained
by dividing $J_{2n}(R)$ by $n J_2S_{2n}(R)/S_2(R)$ for all the
available values $n$ and $R$. The result of such a test is shown in
Fig.2.

\begin{figure}
\epsfxsize=7.6truecm
\caption{ A detailed test of the coefficient $C_{2n}$,
 see text for details. The symbols are the same as in Fig.1. 
The small systematic decrease of $C_{2n}$ with $n$ may be due 
to insufficient accuracy at the tails of the probability density
 function which becomes more important at large values of $n$.}
\label{Fig2}
\end{figure}
\begin{figure}
\epsfxsize=8.3truecm
\caption{The conditional average $H(\Delta T,R)$ measured from the Yale
  data [10] normalized by the measured value of $-J_2/4S_2(R)$ as a
  function of $\Delta T(R)$ for three different values of $R$ measured
  in units of the sampling time. The different $R$ values are
  designated by triangles ($R=16$), squares ($R=128$) and circles
  ($R=1024$) respectively.}
\label{Fig3}
\end{figure}
We see that all the measured values of $C_{2n}$ are
concentrated within the interval $(0.75,1)$ for separations within
the inertial interval. Considering the fact that the quantities
themselves vary in this region over five orders of magnitude, we
interpret this as a good indication for the independence of $C_{2n}$
of $R$ and $n$. The $R$ independence is very clear, and is a direct
test of the fusion rules. The weak $n$ dependence seems to indicate
that $C_{2n}$ decreases slightly with $n$; this may arise from the
limited accuracy of the data. We are reluctant to make a strong claim
about the accuracy of 10'th or 12'th order structure functions.
If we accept for now the evidence that the coefficients $C_{2n}$ 
in (\ref{almostfinal}) are $n$-independent, it must also imply
 that the conditional average $H(\Delta T,R)$ is linear in $\Delta T$. 
 
In Fig.3 we present results from the same data set that was used
above. We show the conditional average $H(\Delta T,R)$ as a function
of $\Delta T(R)$ for various values of $R$. The line passing through
the data points is not a fit, but rather the line required by
Eq.(\ref{linear}). We note that points belonging to different values
of $R$ fall on the same line, indicating that indeed the conditional
average $H$ is a function of $\Delta T(R)$ times a function of $R$,
and that we identified correctly the function of $R$ as $-J_2/ 4
S_2(R)$.

\subsection{Nonlinear Fits}
As explained, the linearity of the conditional average of $H$ in
$\delta T$, [Eq.(\ref{linear})] was not derived from first principles.
We therefore proceed now to see whether the more general form
(\ref{Hform}) is supported by the data, and whether we can bound from
above the values of the parameter $a_1$. To estimate $a_1$ from
experimental data, we first estimate $\alpha(R)$ and $\beta(R)$ from
the pdf's evaluated from data using (\ref{pdf}) then plot $2 H(\Delta
T,R) S_2(R)/[J_2 \Delta T]$ versus $|\Delta T|^{\beta(R)}$. The
intercept is given by $1+3a_1$. To study how well (\ref{Hform}) can
represent the data, we subsitute the estimated value of $a_1$ into
(\ref{Hform}) and compare it with the experimental data. Since the
passive scalar data shown in Fig.3 agree so well with the linear
ansatz, we discuss first a case that offers a more stringent test of
the form (\ref{Hform}). To this aim we consider data taken from
convective turbulence. In this case the temperature is an active
rather than a passive scalar. The data are taken from the well
documented Chicago experiment \cite{87HCL,89SWL}. The experiment was
performed in a cylindrical box of helium gas heated from below, and
the Rayleigh number (Ra) can be as high as $10^{15}$. The box has a
diameter of 20 cm and a height of 40 cm. The temperature at the center
of the box was measured as a function of time, and we use the same
Taylor-hypothesis to surrogate time for the spatial coordinate. Fig.
4  display the conditional average $H(\Delta T,R)$ computed from
these data for three different values of the Rayleigh number Ra with
$R$ measured in units of the sampling time. We see that (\ref{Hform})
is always a good form for describing the experimental data. It is
interesting to examine how the nonlinearity in the conditional average
depends on Ra and on the value of $R$. In Table 1 we present a
compilation of the best fits for the parameters for a range of values
of Ra and for a range of values of $R$. The results appear to support
the following conclusions: 
\begin{itemize}
\item[1.] the value of $a_1$ generally decreases when Ra increases;
\item[2.] the value of $a_1$ is smaller and remains approximately
  constant when the separation $R$ is deep inside the inertial range.
\end{itemize} 

\end{multicols}\widetext
\begin{figure}~
\vskip .7cm
\epsfxsize=18truecm
\caption{The conditional average $H(\Delta T,R)$ as a function of 
$\Delta T$ for the turbulent convection data.
Shown are representative fits of the formula (38) at four values of
 Rayleigh number (Ra) with the separation $R$ measured in units of 
sampling time using Taylor hypothesis.
One sees that the linear fit becomes better as Ra increases, 
and see Table 1 for a quantitative confirmation}
\label{Fig4}
\end{figure}~~
\begin{multicols}{2}
A good way to have a quick estimate of the importance of the nonliear
term [cf Eq.(\ref{Hform})] is to measure $a_1\alpha\beta$ which is the
coefficient of the nonlinear term. We see that this coefficient
decreases significantly when we go from Ra$=6.0\times 10^8$ to
Ra$=5.8\times 10^{14}$, becoming about $0.01$ in the middle of the
inertial range.

Next we show similar detailed calculations for the passive scalar data
of Fig.3. There is a slight complication since the pdf's $P(\Delta
T,R)$ are not symmeteric in $\Delta T$.  Accordingly we need to fit
separately the left and right branches of the distribution functions,
and find the parameters $\alpha_{+}$, $\alpha_{-}$, $\beta_{+}$ and
$\beta_{-}$, together with the appropriate values of $(a_1)_{+}$ and
$(a_1)_{-}$. After doing all this we show the fits in Figs.8-10. It
appears to the eye that the quality of the fits is not significantly
improved compared to the linear fit. To see this more clearly we
present in Table 2 the values of all the parameters involved in the fit.
It is seen that the values of the parameters $a_{1{_\pm}}$ are close
to zero, or more precisely $a_{1_{\pm}}=0\pm 0.02$. The coefficient
that measures the importance of the nonlinear correction, i.e
$a_1\alpha\beta$ is of the order of $10^{-4}$ for all the separations
in the bulk of the inertial range.

\end{multicols}
\begin{table}~\vskip 1cm
\caption{
Fitted parameters for the turbulent convection data,
 with the Rayleigh number spanning the range 
$6\times 10^8$ to $5.8\times 10^{14}$. The parameters $\alpha$
 and $\beta$ characterize the probability density function $P(\Delta
T,R)$,
and $a_1$ is the coefficient of the first nonlinear contribution to the
 expansion of the conditional average $H(\Delta T,R)$ in $\Delta T$. The
 separation $R$ is measured in units of the sampling time. The value of 
the parameter $a_1$ measures the deviation of the coefficients $C_n$
from
 unity, cf. Eq.(28). The combination $a_1\alpha\beta$ measures the
deviation
 of the conditional average from the linear fit at $\Delta T=1$, cf.
Eq.(38).
} ~\vskip .5cm
\begin{tabular}{||c||c|c|c|c|c|c|c|c|c|c||} 
& & & & & & & & & &\\
Ra & $R$ & 8 & 16 & 32 & 64 & 128 & 256 & 512 & 1024 & 2048 
\\ & & & & & & & & & &\\
\hline \hline
& $\beta$ &0.54 &0.64 &0.71 &0.95 &1.06 &1.31 &1.74 &1.84 & --- \\
 $6\times 10^8 $& $100a_1$&$-40.3$&$-22.8$&$-16.5$ &$-12.9$
&$-8$&$-17.6$&$-7.3$&$-16.7$&--- \\ 
& $~100 a_1 \alpha \beta~$ &$~-116~$ &$~~-57~~$ &$~-37.2~$ &$~-24.9~$
 &$~-14.3~$ &$~-14.3~$ &$~-4.4~$ &$~-10.0~$ &--- \\ \hline \hline
& $\beta$ &$0.52$ &$0.67$ &$0.85$ &$1.05$ &$1.19$ &$1.41$ &$1.81$
&$2.00$
 &--- \\
$4\times 10^9 $ & $100a_1$ &$-24.8$ &$-14.6$ &$-12.8$ &$-5.4$ &$-7.2$
&$-6.3$
 &$-3.1$ &$-16.4$ &--- \\
& $100a_1 \alpha \beta$ &$-68$ &$-33.4$ &$-22.4$ &$-8.7$ &$-9.0$ &$-8.1$ 
& $-2.6$ &$-10.0$ &--- \\
\hline \hline
 & $\beta$ &--- &$0.61$ &$0.68$ &$0.80$ &$0.91$ &$1.15$ &$1.38$ &$1.55$ 
&$1.68$ \\
$ 7.3\times 10^{10} $ & $100a_1$ &--- &$-9.1$ &$-7.3$ &$-7.1$ &$-1.1$
&$-16.3$ 
&$-12.1$ &$-14.8$ &$-5.7$ \\
& $100a_1 \alpha \beta$ &--- &$-24.5$ &$-17.6$ &$-12.1$ &$-3.3$ &$-15.5$
 & $-9.5$ &$-11.1$ &$-3.0$ \\
\hline \hline
 & $\beta$ &--- &$0.60$ &$0.65$ &$0.72$ &$0.84$
 &$0.96$ &$1.19$ &$1.50$ &$1.43$ \\
$6\times 10^{11}$& $100a_1$ &--- &$-6.9$ &$-2.8$ 
&$-14.0$ &$-12.9$ &$-10.9$ & $-13.7$ &$-4.4$ &$-14.2$ \\
& $100a_1 \alpha \beta$ &--- &$-18.8$ &$-9.1$ &$-27.1$ 
&$-20.6$ &$-71.0$ & $-15.7$ &$-4.7$ &$~-11.6~$ \\
\hline \hline
 & $\beta$ &--- &$0.58$ &$0.64$ &$0.72$ &$0.83$ &$0.94$ &$1.00$ 
&$1.25$ &$1.45$ \\
$6.7\times 10^{12} $& $100a_1$ &--- &$-7.1$ &$-20.4$ &$-17.5$ 
&$-8.2$ &$-6.7$ & $-3.9$ &$-7.1$ &$-8.1$ \\
& $100a_1 \alpha \beta$ &--- &$-22.9$ &$-48.1$ &$-34.9$ &$-14.4$ 
&$-12.4$ & $-9.6$ &$-9.5$ &$-7.6$ \\
\hline \hline
 & $\beta$ &--- &$0.61$ &$0.68$ &$0.77$ &$0.86$ &$0.88$ &$1.15$ &$1.26$ 
&$1.42$ \\
$4.1\times 10^{13} $& $100a_1$ &--- &$-2.9$ &$-1.5$ &$-0.5$ &$-2.3$ & 
$-1.2$ &$-0.7$ &$-30.7$ &$-10.5$ \\
& $100a_1 \alpha \beta$ &--- &$6.4$ &$-4.7$ &$-1.9$ &$-6.0$ &$-2.4$ 
& $-2.4$ &$-22.5$ &$-9.4$ \\
\hline \hline
 & $\beta$ &--- &$0.59$ &$0.63$ &$0.69$ &$0.85$ &$0.93$ &$1.05$ &$1.31$
 &$1.56$ \\
$~5.8\times 10^{14}~ $& $100a_1$ &--- &$-4.7$ &$-2.6$ &$-4.5$ &$-5.3$ & 
$-0.4$ &$-2.5$ &$-8.7$ &$-3.1$ \\
& $100a_1 \alpha \beta$ &--- &$-10.9$ &$-4.7$ &$-6.0$ &$-4.6$ &$-1.0$
 &$-4.3$ &
$-8.9$ &$-2.8$ \\
\end{tabular}\end{table}~\vskip 1cm
\begin{table}\caption{The fit parameters for the nonlinear fits 
in the passive advection data with Reynolds number (Re) $=9.5\times
10^4$.
 The probability
density function $P(\Delta T,R)$
is asymmetric in this case, and we fit separately the left (minus
subscript)
 and right (plus subscript) wings. All the parameters
that measure the deviation from the linear fits are very small. } ~
\vskip .5 cm
\begin{tabular}{||c|c|c|c|c|c|c|c|c||}
& & & & & & & &\\
$R$ & 8 & 16 & 32 & 64 & 128 & 256 & 512 & 1024 \\ & & & & & & & & \\
\hline \hline
$\beta_-$ &0.94 &1.12 &1.28 &1.54 &1.77 &1.85 &1.76 &1.86 \\ $\beta_+$ 
&1.15 &1.32 &1.56 &1.63 &1.75 &1.82 &1.87 &1.82 \\ \hline $10^3 (a_1)_-$ 
&21 &67 &11 &12 &$-3$ & 9 &10 &$-5$ \\ $10^3 (a_1)_+$ &18 &14 &$-1$ &19
 &$-6$ &6 &0.6 &12 \\ \hline 
$ 10^3(a_1 \alpha \beta)_-$ &10 &16 &3 &1 &$-0.4$ &0. &0.2 &$-0.4$ \\
 $10^3 (a_1 \alpha \beta)_+$ &1.8&$-0.7$ &$-1.7$ &0.2 &$-0.3$ 
&$-0.1$ &$-0.2$
& 0.4 \\
\end{tabular}
\end{table}

\begin{figure}~~\vskip 1cm
\epsfxsize=11truecm
\caption{
The conditional average $H(\Delta T, R)$ as a function of $\Delta T$ 
for the passive scalar data at three values of the separation $R$
measured 
in units of sampling time. The nonlinear fits are indistinguishable from 
the linear ones in the bulk of the
inertial range, and see Table 2 for a quantitative confirmation}
\label{Fig5}
\end{figure}
\begin{multicols}{2}
\section{conclusions}
We have presented a theoretical analysis of the relation between the
two conditional averages $H(\Delta T,R)$ and $G(\Delta T,R)$ and the
probability distribution function $P(\Delta T,R)$. The general
relation is given by Eq.(\ref{univ}). From this relation it follows
that if one of these averages factorizes to a function of $\Delta T$
times a function of $R$, the other cannot factorize as long as the the
distribution function does not factorize. The latter cannot factorize
in any multiscaling statistics. Next we have presented evidence that
the conditional average $H(\Delta T,R)$ does factorize into a function
of $R$ times $\Delta T$. This appears to be the case for both passive
and active scalars when the Reynolds number is sufficiently large and
when $R$ is in the bulk of the inertial range. The fusion rules which
are believed to hold in a wide context furnish a prediction about the
function of $R$ that precedes $\Delta T$ in the conditional average,
cf. Eq.(\ref{linear}). The data support the prediction of the fusion
rules to very high accuracy. We do not have at present a theoretical
ab-initio derivation of the linear dependence that seems to be
supported by the data. In view of the importance of this law for the
study of the balance equation $D_n(R)=J_n(R)$ it seems to us that such
a derivation is highly desirable.

\acknowledgments This work was supported in part by the German Israeli
Foundation, the US-Israel Bi-National Science Foundation, the Minerva
Center for Nonlinear Physics, and the Naftali and Anna
Backenroth-Bronicki Fund for Research in Chaos and Complexity. The
work of ESCC is also supported by the Hong Kong Research Grants
Council (Grant No. 458/95P).

\end{multicols}
\end{document}